\documentclass[journal]{IEEEtran}

\usepackage{xcolor,soul,framed} 
\usepackage{booktabs}
\colorlet{shadecolor}{yellow}
\usepackage[pdftex]{graphicx}
\graphicspath{{../pdf/}{../jpeg/}}
\DeclareGraphicsExtensions{.pdf,.jpeg,.png}
\usepackage{graphics}
\usepackage[cmex10]{amsmath}
\usepackage{array}
\usepackage{mdwmath}
\usepackage{mdwtab}
\usepackage{eqparbox}
\usepackage{url}

\usepackage{cite}
\usepackage{vcell}

\usepackage[numbers,sort&compress]{natbib}

\newcommand{\subheading}[1]{\noindent{\textbf{#1}}}

\begin{document}
\bstctlcite{IEEEexample:BSTcontrol}
    \title{Cybersecurity Challenges of Power Transformers}
  \author{Hossein~Rahimpour,~Joe~Tusek,~Alsharif~Abuadbba,~Aruna~Seneviratne,~\IEEEmembership{Senior Member,~IEEE}\\~Toan~Phung,~\IEEEmembership{Senior Member,~IEEE,},~Ahmed~Musleh,~\IEEEmembership{Member,~IEEE,}\\~and~Boyu~Liu,~\IEEEmembership{Member,~IEEE}\\

  \thanks{This work was funded by Australian Government Research Training Program, and in part by cybersecusity Cooperative Research Centre through CSIRO's Data61.}}



\maketitle

\begin{abstract}
The rise of cyber threats on critical infrastructure and its potential for devastating consequences, has significantly increased. The dependency of new power grid technology on information, data analytic and communication systems make the entire electricity network vulnerable to cyber threats. Power transformers play a critical role within the power grid and are now commonly enhanced through factory add-ons or intelligent monitoring systems added later to improve the condition monitoring of critical and long lead time assets such as transformers. However, the increased connectivity of those power transformers opens the door to more cyber attacks. Therefore, the need to detect and prevent cyber threats is becoming critical. The first step towards that would be a deeper understanding of the potential cyber-attacks landscape against power transformers. Much of the existing literature pays attention to smart equipment within electricity distribution networks, and most methods proposed are based on model-based detection algorithms. Moreover, only a few of these works address the security vulnerabilities of power elements, especially transformers within the transmission network. To the best of our knowledge,  there is no  study in the literature that systematically investigate the cybersecurity challenges against the newly emerged smart transformers. 
This paper addresses this shortcoming by exploring the vulnerabilities and the attack vectors of power transformers within electricity networks, the possible attack scenarios and the risks associated with these attacks.

\end{abstract}




%
\IEEEpeerreviewmaketitle


\section{Introduction}   

\IEEEPARstart{D}{igitalisation} and automation are transforming the power systems. The transformation is being facilitated through capabilities of power systems being able to access and communicate via communications networks (network connectivity) as well as the introduction of Artificial Intelligence (AI) enabled diagnostic systems \cite{NakahataDigitalizationMaintenance,HONG2014CYBERSYSTEMS}. Prior to 2010, there were very few  connections between Operational Technology (OT) and the Information Technology (IT) \cite{Gregory-Brown2016SecurityWorld}. The reviews of the advanced computer aided communication systems in the smart grid context show that these systems add significant value to the productivity and efficiency of power systems \cite{Pour2017ASystems,Shrestha2020AInfrastructures}. 

Until recently, there were only a few connections and very little knowledge about the OT infrastructure of power systems. Hence it was not on the radar of the cybersecurity, and broadly, the industry had few if any security measures apart from physical isolation in place. Over the last decade, the OT and IT domains have started to merge through the introduction of advanced monitoring and control systems, as well as remote access and control systems. However,  emerging cybersecurity challenges are the downside of these new technologies as it has been discussed in the literature \cite{Marksteiner2019CyberModeling,Otuoze2018SmartThreats,Sun2018CyberState-of-the-art,Reich2020CybersecurityProviders}.

With the integration of IT with OT infrastructures, there is a need to consider the cyber risks, identify the gaps in security frameworks \cite{CIGRED2.382017FrameworkInfrastructure,Sarkar2022AAutomation}, policies, and introduce effective mitigation strategies \cite{CIGREWGB5.662020CIGRE2020.,CIGREWGD2.462020CIGRE2020.,Devanarayana2019TestingSimulator,Taljaard2018NetworkD2-309,Dondossola2020CyberSystems} as the OT becomes exposed to cyber threats \cite{CenterforStrategicandInternationalStudies2021Significant2006}. 


A power transformer is a critical end device asset within the electricity network that is undergoing digitalisation, as it is fundamental to the production and transmission of power to customers \cite{Shi2020AServices}.  Thus, power transformers could be targeted to potentially induce catastrophic failure of the power systems and significant and long-term supply disruptions \cite{9372271,Weiss2019LargeYears,Koelemij2020OTTransformers.pdf}.
The digitisation of transformers happens by inclusion of online condition monitoring (OLCM) \cite{Ivar-Ulvestad-Raanaa2020ConditionSubstations,ElectricPowerResearchInstituteEPRI2013IECData,Dolata2011On-Line61850,Hui2012TheTransformer} equipment with AI enabled diagnostic systems. The real time monitoring equipment using data acquisition systems produces enormous amount of data that needs to be protected. The data either is used to indirectly operate a critical device or influence a decision to be made at supervisory level of a substation.
Data security is becoming the main focus of the power transformer OT security. There are many types of attacks including network topology attack, False Data Injection (FDI), jamming attack, GPS spoofing \cite{Wang2018DistributedClocks}, and time synchronization attack\cite{Shereen2020FeasibilityEstimation}. However, among all these atttacks FDI has been found the most known common attack \cite{Lu2022GeneticSystems,Unsal2021EnhancingMitigation,Yang2017TowardGrid,Hittini2020FDIPP:Systems,Dayaratne2022FalseGrids,Mujeeb-Ahmed2018NoisePrint:Systems}. 
This even becomes more critical for assets such as power transformers that are equipped with diagnostic data acquisition systems with a life span of 40 years in the electrical network. The associated hardware and software will also be required to match that the 40 years specification. When comparing the existing IT cybersecurity technologies to the ones that are required in the power industry as part of Operational Technology (OT), there are major differences. The traditional network defense mechanisms for cyber elements in power system application have been challenged by researchers \cite{Karimipour2019AGrids}. The long life span of power system equipment, the strict timing requirements, higher availability, less latency requirements, mostly asymmetric and message oriented communication technologies are only a few differences that make the power networks unique in comparison to IT technologies of the office networks. These have been highlighted in IEC standards \cite{2022IEC/TRGuidelines} and literature \cite{Maglaras2018CyberInfrastructures,Taljaard2018NetworkD2-309,Dayaratne2022FalseGrids}.

Although there are numerous studies on various aspects of smart grid cybersecurity that highlight the criticality of the cybersecurity in smart grid technologies \cite{Shrestha2020AInfrastructures,Sun2018CyberState-of-the-art,Kumar2019SmartIssues,Sakhnini2021SecuritySurvey,Baumeister2010LiteratureSecurity} and intelligent grid equipment such as smart meters \cite{Wang2019ReviewChallenges,Hassan2022AMeters}, to the best of our knowledge there are no studies  that focus on smart transformers. In this paper, we address this gap and make the following contributions:    
1)	Providing smart transformer's architecture and well-formed taxonomy of the relevant cyber physical attacks. 
2)	Detailed review of smart transformers topology, vulnerable components and the implication of cybersecurity breaches. 
3)	Discussion of the potential future cybersecurity challenges of smart transformers. 

The remainder of the paper is orgnaised as follows. Section II introduces the smart transformers topology and its emerging condition monitoring technologies. Then, it details intelligent transformers and their individual components in relation to the substation architecture. Finally, it presents phase shifting transformers. Section III summarises some of the major historical attacks on critical infrastructure in the world. Section IV discusses the vulnerabilities and attack vectors specific to smart grids and section V details the attacks specific to power transformers. Anomaly detection algorithms and a summary of the simulation tools and test beds are presented in sections VI and VII respectively. The use of AI techniques for cybersecurity of power transformers has been discussed in section VIII and finally the conclusion is given in section IX. 







 
  
\section{Smart Transformers Topology}


Electric power is generated by converting other forms of energy into electricity. The power produced is then transmitted geographically, typically over long distances using transmission lines, through transmission substations and then through distribution substations and ultimately to consumers. Transformers play a very important role in matching voltage levels to the needs of the network and to those of its customers. Arguably the most concentrated, expensive, and critical components of the system, transformers are long-lived assets with a service life of greater than 40 years. As they are one of the most expensive elements of the grid, their integrity and operational performance are always of concern to utilities. Thus, technologies that can assist in their efficient use and early diagnosing of degradation monitoring are valuable within the asset management frameworks as they are used to gain the greatest value for the installed equipment.

Even so, it is observed that for power transformers, the adoption of digitisation has been slow and it is only since around 2018 that major players in the industry have been launching digital smart transformer products. For example, Reinhausen (also known as Maschinenfabrik Reinhausen MR) launched ETOS® - Embedded Transformer Operation System for operation, control and monitoring of transformers \cite{MaschinenfabrikReinhausenGmbHETOSSystem}. Siemens Energy, has likewise launched Internet of Things (IOT) equipped transformers as "Sensformer" \cite{SiemensEnergySensformerSecurity} and Hitachi Energy has developed TXpert™ Enabled digital power transformers \cite{ABBsHitachiEnergy2018TXpertTransformer}. 
All these companies highlighted the potential cybersecurity challenges and they are working towards solutions to overcome existing and future cybersecurity issues associated with the new generation of  transformers. 

Typically such transformers are known as smart/digital transformers and are conventional transformers equipped with intelligent monitoring systems, aimed at the  early detection of abnormalities within the transformer.  Due to the gains in operational and financial efficiency afforded by such systems, it is anticipated that there will be a wider adoption of such technology across existing transformer fleets. 
While such systems provide considerable benefits, they do open the door to new vulnerabilities, via connections through data communication channels. 
This includes the substation client server, real time services and the GPS communication services. In a worst-case scenario for the network, it may be possible for a cyber attack to gain control over switching operations leading to local catastrophic consequences and potential long term supply disruption. In a lesser case, the sending of false instrument reading may make the operator take them offline otherwise healthy equipment leading to supply restrictions or blackouts by an incorrect decision.  
To date, there has been little attention paid to the potential cyber threats against power transformers \cite{8274710,10.1007/978-3-319-23802-9_20,8909786,9126843,Jahromi2020,9137377,Olijnyk2022DesignTransformer,Ahmad2022AdvancedSubstation}. Also very few writers have been actively warning of the emerging cyber-attack issues for transformers \cite{Koelemij2020OTTransformers.pdf,GoudCyberExplosions,King2021HowCyberattack,Ribeiro2021ChineseAdministrations}. Digitisation of transformers involves a suite of sensors, data collection and analysis tools to be installed on every transformer to allow for real-time monitoring (Figure 1). With digitisation, vulnerabilities of these new technologies to cybersecurity is an ever-present threat. 




\begin{figure}[h]
    \includegraphics[width=\linewidth]{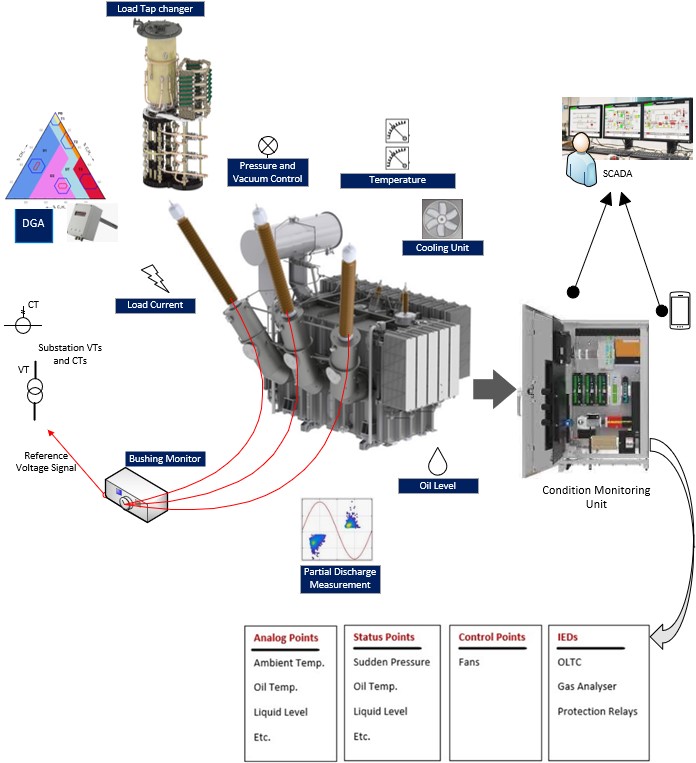}


  \caption{Power transformer monitoring system } \label{Monitoring}
\end{figure}

The following sections will be detailing the transformer within substation architecture and how the transformer itself and its condition monitoring devices are connected to the rest of the network. Then each smart component of a transformer has been closely investigated to understand the involved communication protocols and data transmission technologies. 

  


\subsection {Substation Architecture and Transformers}
Over the years conventional substations are transformed into modern and digital substations, through replacement of copper wired connections with advanced communication technologies. This is driven by the need to achieve higher efficiency and to lower the cost of operation. However, with multi-vendor systems and their associated increased connectivity, there is also the evolution of unprecedented vulnerabilities \cite{JACOBS2021PowerControl,MissionSupportCenterIdahoNationalLaboratory2016CyberSector,Sahoo2021CyberVulnerabilities,Hahn2013CyberEvaluation}. This exposes the efficiency and stability of the network to cyber risks with significant operational and financial consequences \cite{Musleh2020}.

Various physical critical equipment is located in substation architecture as it is shown in Figure 3 \cite{OmicronDetectingNetworks}. 
Typically this architecture is structured in the following three levels:

1)	Station level: Supervisory systems, Local Supervisory control and data acquisition (SCADA) systems and substation automation.

2)	Bay level: Control functions and System protection 

3)	Process level: Device level

Transformers and other substation field devices such as Circuit Breakers (CB), Current Transformers (CT) and Voltage Transformers (VT), etc., are positioned at  the process level. The analogue signals produced by these field devices get converted to digital signals through merging units. The bay level equipment comprises of the control, protection and measurement of Intelligent Electronic Device (IED). The station level includes the elements that provide a human-machine interface (HMI), dealing with the station computers and gateways.   
The risk modelling performed by a South Australian utility \cite{Automated2020ParisGrid} shows that process level data information has the highest risk within a substation environment. This shows the necessity of having a defence in the process level equipment and communication channels associated with these devices.
Conventional substations hard wired the station level systems to bay level devices and to process level equipment using copper conductors. Later with the need for improvement in network reliability and stability, rapid fault isolation and restoration \cite{Kumar2021Toward61850}, the modern substations emerged, replacing the hard-wired connections and serial connections with data network communication protocols.  
Three types of communication are available in the substation environment:

1) Distributed Network Protocol (DNP) 3.0 or IEC 60870-5; at the station level, the data exchange in and out of substation (servers, HMIs and gateways, including the remote access).

2) Generic Object Oriented Substation events (GOOSE) messages; data exchange between relays and protection devices and interlocking.

3) Manufacturing Message Specification (MMS); used for Control and protection of data exchange between station and bay level.

Conventional systems are still using Modbus and DNP3 which are considered vulnerable to cybersecurity \cite{10.1007/978-3-319-23802-9_20}. Despite the use of modern IEC 61850 protocols for substations, this communication protocol has also proven to be vulnerable to attacks.       
The purpose of using such communication protocols is to enable the transfer of the process level and bay level equipment information to network supervisory systems such as SCADA and Energy Management Systems (EMS). This enables grid operators to make informed decisions about the state of the energy supply system and to facilitate the reliable and secure supply of electricity to network customers \cite{CEN/CENELEC/ETSIJointWorkingGrouponStandardsforSmartGrids2012CEN-CENELEC-ETSISecurity}.
With the transition to digital substations, the switchyard primary equipment such as transformers with a fast-transient response has been given special attention by International Electrotechnical Commission 61850 due to their critical role in the network \cite{Kumar2021Toward61850}.

\subsection {Transformer condition monitoring equipment- Device level}

In the current market, there is an ongoing effort toward the development and advancement of sensors, monitoring tools and digital interfaces for transformers. These are to support the transformer fleet at three levels: generation, transmission, and distribution with modular integration of functions in the following areas: 

1.	Voltage regulation using Onload Tap-Changers (OLTC): Drive, Control and monitoring,

2.	Transformer insulation and bushing condition, 

3.	Transformer internal temperature, and control of fans and pumps,

4.	Dissolved Gas Analysis (DGA) system; Gases produced by discharge activity inside the transformer tank.

The above-mentioned monitoring and diagnostics devices could be tampered with and pose a hazard to the operation of transformers and the power network. False alarm signalling, tampering with temperature indication devices, manipulation of OLTCs in setting voltage levels. Maloperation of fans by manipulating temperature gauges is also a way of causing overheating in transformers. As the loss of a transformer in operation will lead to costly outages, loss of production, and possible widespread blackouts, it is important that the integrity of the controls and indications is maintained.


\subsection {Transformer OLTC}

OLTCs are an important component for the regulation of voltages on the network and directly influence the network's voltage stability. If the voltage stability is targeted or impacted negatively during a peak load condition, it can result in voltage collapse, failures and consequently blackouts. OLTC control systems are in use for transmission transformers to ensure dynamic and adaptive control of voltages. These intelligent devices increase the attack surface as the control signals travel through the communication channels of the network. Any malicious data modification or manipulation of voltage control on OLTCs could result in a maloperation \cite{10.1007/978-3-319-23802-9_20}. 
OLTCs are one of the vulnerable parts of the transformer likely to be targeted by attackers who gain access through the RTU or SCADA. Transmission network operators change the operating tap of transformers for many reasons and by various means. This includes the utilization of solutions from load flow and optimal power flow (OPF) calculations \cite{9126843}. 
Vulnerabilities of OLTC have been analysed for two types of attacks \cite{9126843,10.1007/978-3-319-23802-9_20}. This includes False Data Injection (FDI) and stealthy false command injection attacks. As the FDI has been found to be the most common attack in studies on transmission systems, the false control signal injection can be even more hazardous when the attack is performed stealthily with the aid of FDI \cite{9126843}. A. Anwar \cite{10.1007/978-3-319-23802-9_20} has addressed the risks and the impacts of FDI attack on OLTC with two attack scenarios when more power is demanded in the system, and when nodes of the systems are placed under the lower stability margin, however, no mitigation strategies have been provided.  
S. Chakrabarty \cite{9126843} has applied an estimation-based algorithm with a specific threshold that provides a 100\% success rate. However, this may not be an effective mitigation strategy for the future growing grid and increased complexities, as well as the advancement of attacks. The application of model-free data-driven techniques capable of detection with lower thresholds can be an effective defence mechanism for cyber securing OLTCs.  


\subsection {Transformer insulation and its bushings condition}

The monitoring tools are designed to communicate the transformer health data to SCADA or Digital Control System (DCS) through relays for warning, alarm and error status. 
The active part of transformers has a higher failure rate and the monitoring of the windings, insulation condition as well as bushings are thus of paramount importance. These new monitoring systems are designed to provide information on the status of the transformer insulation including bushing condition through several indicators. This information is critical to the proper assessment of the transformer’s health and is dependent on the known condition and the remaining life of its insulation. The paper and oil insulation used in power transformers can tolerate the high temperatures under normal operating conditions. However, if the transformer is overloaded for an extended period, over-fluxed, or has its cooling interrupted, then the insulation temperature may exceed design limits, greatly accelerate the decomposition of its paper and lead to the degradation of the insulation reducing its electric strength and mechanical integrity. If this is unchecked or misleading information is sent to supervisory systems because of an attack, the ultimate result would be a devastating failure or a major disruption to the power supply.  
Typically, the insulation parameters of Partial Discharge (PD), Dielectric Dissipation Factor (DDF) and Capacitance are measured by the monitoring system. To the best of authors knowledge this attack vector has not been identified or discussed in the existing literature. 



\subsection {Transformer internal temperature, and control of fans and pumps}

Transformer life expectancy is very dependent on its operating temperature and more specifically on the hot spot temperature in the transformer \cite{Heathcote2007TheBook}. To understand the importance of this, Australian Standard AS60076 indicates that every 6 degrees increase above the allowed temperature, will result in halving of the expected transformer life. Thus, on large power transformers, there are typically fans and pumps used with the addition of large radiators.




To control the internal temperature, monitoring systems send temperature indications to the supervisory systems so that fans and pumps can be operated. Any FDI via cyber-attack can result in the maloperation of fans and pumps and possible overheating and failure of the transformer.



\subsection {Transformer DGA System}

DGA is one of the key transformer monitoring tools that can be used in both online and offline service conditions. This tool measures the gases accumulated in transformer oil and can be used to infer faults or stresses within the insulation of the transformer \cite{Brochure2019D1/A2Systems}. Receiving any false data or control commands through DGA units could lead to legitimate stress indications being  manipulated into benign readings. As a result,  incorrect decision-making of the supervisory system could  cause consequences from a possible unnecessary outage to a failure scenario at worst. B. Ahn \cite{9372271} has explored the vulnerabilities of two commercial DGA systems and discussed the potential threats using a threat model. The threat has been identified at three levels of 1)	 Device Level, 2)	Digital Substation, and 3)	Remote Access. 

At the device  level, the USB slave connection of the device could be the vulnerable point to malware injection or unauthorised access. As a result of such an attack, the firmware can be manipulated, and the data storage could also be targeted. The windows-based program of these systems is connected through an Ethernet connection supported by DNP3, Modbus and IEC61850. Attacks that bypass the firewall can break down the authentication by Structured Query Language (SQL) injection. FDI and Denial of Service (DoS) are the examples of  attacks at the substation level. Remote access level also provides a back door for attacks such as Malware injection, Spoofing and DoS.



\subsection {Transformer associated devices - substation level}

Transformers can be attacked through their associated equipment at the substation level. The overall grid performance can be significantly impacted by the disruption of these devices by attackers. Based on \cite{8274710,Jahromi2020} the following devices associated with transformers can be tampered with and lead to failures and outages.  
1.	Metering equipment; Current Transformers (CT) and Voltage Transformers (VT)   
2.	Circuit Breakers (CB)
3.	Differential protection 
The involvement of industrial control systems (ICS) with the existing electrical process and mechanical functions, along with connectivity to the internet, creates another attack surface in the network.
CTs and VTs, as the substation metering devices collect the electrical data from the power transformers as an analogue signal and feed them to the merging units (MU). The MUs then convert those to digital data packets and transfer them via switches and IEDs to control rooms \cite{Kumar2021Toward61850}. 
The data packets are then carried as Sampled Values (SVs) and Generic Object Oriented Substation events (GOOSE) messages to control systems. Any change to the size of their byte size will result in sending a command to circuit breakers to interrupt the power flow to protect the equipment.
The state of the circuit breakers (open/close conditions) can now be managed remotely. This brings efficiency and cost savings for utilities, however, it provides more access to adversaries \cite{SenateRPC2021INFRASTRUCTUREGRID}.  Targeted periodic overloading can lead to degradation of the insulation of transformers over time and ultimately to failure and loss of operation. Hypothetical scenarios of the periodic overloading of a transformer bank (3 x 1 phase transformers) have been studied and the impacts have been shown with simulations \cite{8274710}.
Papers \cite{Jahromi2020,Olijnyk2022DesignTransformer,Khaw2021ARelays} have studied the cyber-attack scenarios on differential protection devices associated with transformers as a critical asset in the power network. Disruption to such protection equipment can lead to catastrophic failures at the network operation level. Detection of the abnormal behaviour of protection relays because of cyber threats is becoming very important for utilities.

\subsection {Transformer remote control systems – remote access}

As the condition monitoring data will be accessible through engineering PCs, workstations, and cloud based remote systems for analysis, this creates multiple points of access for adversaries.
The sensitive information that is collected through merging units (MU), Remote Terminal Units (RTU) and SCADA is often remotely accessible for further analysis by engineering staff. This increases the chance of attacks through these vulnerable points in the network. 
The use of remote interfaces has even been further encouraged by service providers during the Covid pandemic, as a means of coping with reduced site access.


\subsection {Special transformers; Phase Shifting Transformers}

Phase shifting transformers (PST) are used to control the active power flow in the power grid, prevent overloading and cross-network regulation of power flow. As their commands are typically transferred through SCADA systems, they can be readily targeted by cyber-attackers. 
The consequences of such an attack could be transformer overloading and outages of transmission lines as well as cross-network trading losses \cite{9137377}. 
PST are controlling the real power flow by either enforcing or blocking it by regulating the phase displacement as these two are very closely connected. The phasing shifts (either leading or lagging) by changing the tap position across its tapping range. Remote Terminal Units (RTUs)  facilitate the connection of PSTs, through SCADA to the command centre \cite{9137377}. 
Paper \cite{9137377} explored attack scenarios targeting phase shifting transformers by gaining access to the RTUs and proposed a detection algorithm for discerning unexpected phase shifts. Malicious phase shift command attack has been proved to be a potential cause of generation load imbalance and in severe cases, lead to significant financial losses.


\hspace{1cm} 


\section{Simulation Tools and Test Beds}

To achieve a comprehensive level of analysis across a complex and fast changing power grid, modelling and simulation are important. Papers \cite{Yohanandhan2020Cyber-PhysicalApplications,Smadi2021AChallenges} provide a survey of simulation platforms and available test beds. Often test beds have been designed for specific tasks or research fields and they provide limited set scenarios. However, some provide greater flexibility in both cyber/ software platforms as well as the physical hardware elements.  Table 2 also provides a summary of the most used modelling and simulation tools.  Paper \cite{Devanarayana2019TestingSimulator} demonstrates the importance of the simulation environment in developing mitigation strategies. 
Real-time simulators have improved the opportunities for large scale grid simulation and allow for coverage of dynamic features of grid equipment, such as inverters \cite{Song2021ResearchGrid}.  Models can be applied to different infrastructures; electrical models, communications/cyber system models, and co-simulation models. 
For transformers, thermal modelling also has significant importance as overloading \cite{8274710} has been shown to be one of the most possible scenarios from cyber attacks.

\begin{table}
\centering
\caption{List of simulation software used in literature}
\begin{tabular}{|l|l|l|l} 
\cline{1-3}
\textbf{Power System~} & \textbf{Cyber System~} & \textbf{Co-Simulation~} &   \\ 
\cline{1-3}
RTDS-RSCAD~        & ns2 and ns-3           & ParaGrid                &   \\ 
\cline{1-3}
MATLAB-Simpower~       & OMNet++                & Modelica                &   \\ 
\cline{1-3}
GridDyn                & Java                   & Dymola                  &   \\ 
\cline{1-3}
PSCAD/EMTDC            & RINSE                  & MathModelica            &   \\ 
\cline{1-3}
PowerWorld Simulator   & OPNET                  & MapleSim                &   \\ 
\cline{1-3}
OpenDSS                & Visual Studio          & JModelica               &   \\ 
\cline{1-3}
DIgSILENT              & GridSim                & Ptolemy II              &   \\ 
\cline{1-3}
EMTP-RV                & NeSSi2                 & Simantics               &   \\ 
\cline{1-3}
OPAL-RT                & GridStat               & Mosaik                  &   \\ 
\cline{1-3}
ETAP                   & DeterLab               & Simscape                &   \\ 
\cline{1-3}
GridLab-D              & WANE                   & EPOCHS                  &   \\ 
\cline{1-3}
PSLF                   & UPPAAL                 & Simulink                &   \\ 
\cline{1-3}
MATPOWER               & Stateflow              & LabVIEW                 &   \\ 
\cline{1-3}
EnergyPlus             & TIMES-Pro              &                         &   \\ 
\cline{1-3}
PowerFactory           & MATLAB - SimEvents     &                         &   \\ 
\cline{1-3}
UWPFLOW                & GLOMOSIM               &                         &   \\ 
\cline{1-3}
TEFTS                  & Cloonix                &                         &   \\ 
\cline{1-3}
PST                    & GNS3                   &                         &   \\ 
\cline{1-3}
InterPSS               & IMUNES                 &                         &   \\ 
\cline{1-3}
OpenPMU                & Shadow                 &                         &   \\ 
\cline{1-3}
rapid61850             &     EXata                   &                         &   \\ 
\cline{1-3}
Aspen                  &                        &                         &   \\ 
\cline{1-3}
PLECS                  &                        &                         &   \\ 
\cline{1-3}
adevs                  &                        &                         &   \\ 
\cline{1-3}
NEPLAN                 &                        &                         &   \\ 
\cline{1-3}
EUROSTAG               &                        &                         &   \\ 
\cline{1-3}
Homer                  &                        &                         &   \\ 
\cline{1-3}
PCFLO                  &                        &                         &   \\ 
\cline{1-3}
Psap                   &                        &                         &   \\
\cline{1-3}
\end{tabular}
\end{table}

\section {Historical cybersecurity in Electrical Network}

\subheading{Attacks to Australian Critical Infrastructure.}
According to the Australian cybersecurity Centre (ACSC) report, 67,500 cybercrimes with an estimated loss of \$33 billion were reported in the financial year 2020-2021. This is an almost 13\% increase when compared to the previous year 2019-2020. Almost one quarter of these attacks affected critical infrastructure which includes electricity networks \cite{AustralianCyberThreatCentre2021ACSC2020-21}. There is no doubt that cyber incidents are on the rise and the electricity network is being targeted. A recent ransomware incident has been reported by CS Energy the Australian based generation company on 27th Nov 2021 \cite{CSEnergy2021CSINCIDENT}.


\subheading{Worldwide Attacks.}
US based utilities ranked the risk of cybersecurity as 4.37 out of 5. The risk studies of UK's Cambridge centre estimate the losses over the 5 years as GPB 442 billion \cite{TransGrid2018TransGridConditions}. Among reported attacks, energy sector attacks have been at the top of the list for the USA (CIGRE Technical Brochure 833 \cite{JACOBS2021PowerControl}).
Utilities worldwide have reported that the level of sophistication and the frequency of cybersecurity has continued to increase \cite{MissionSupportCenterIdahoNationalLaboratory2016CyberSector}. In 2010, Iranian’s nuclear plant PLCs were infected through engineering PCs. In 2015 Ukrainian power plant control centre PCs were remotely controlled and RTUs infected and destroyed. In 2016, a Ukrainian power plant was attacked by malware. In 2017 PLC malware targeted plant safety systems in the Middle East. Table 1 shows the history of some of the major reported cybersecurity in relation to the electrical network and other critical infrastructure. 
One of the major historical attacks on the electricity network is the remote cyber attack directed against Ukraine’s electricity infrastructure in Dec 2015 which maliciously operated SCADA systems. This caused a blackout that hit part of the Ukrainian’s capital city and affected approximately 80,000 customers \cite{Mohan2020ASystems}.










\begin{table*}[t]
\centering
\caption{Significant Attacks History}
\begin{tabular}{|l|l|l|l|l|l|l|} 
\hline
\vcell{\begin{tabular}[b]{@{}l@{}}\\\textbf{Year}\end{tabular}} & \vcell{\textbf{Location}}   & \vcell{\textbf{Attack}} & \vcell{\textbf{Type}}     & \vcell{\textbf{Targeted/ Infected}}                                                                               & \vcell{\textbf{Consequence}}                                              & \vcell{\textbf{Infrastructure}}  \\[-\rowheight]
\printcelltop                                                   & \printcelltop               & \printcelltop           & \printcelltop             & \printcelltop                                                                                                     & \printcelltop                                                             & \printcelltop                    \\ 
\hline
\vcell{2021}                                                    & \vcell{USA}                 & \vcell{Ransomware}      & \vcell{Malware}                  & \vcell{Colonial pipeline}                                                                                         & \vcell{\begin{tabular}[b]{@{}l@{}} Energy company shut down the\\ pipeline and paid a 5MM \$ ransom\end{tabular}}    & \vcell{Oil}                         \\[-\rowheight]
\printcelltop                                                   & \printcelltop               & \printcelltop           & \printcelltop             & \printcelltop                                                                                                     & \printcelltop                                                             & \printcelltop                    \\ 
\hline
\vcell{2021}                                                    & \vcell{Florida, USA}        & \vcell{\begin{tabular}[b]{@{}l@{}} Unauthorized \\Access\end{tabular}}                & \vcell{Stolen Credentials
}                  & \vcell{Water treatment plant}                                                       & \vcell{\begin{tabular}[b]{@{}l@{}}Boosted treatment chemicals\\ (NaOH) to dangerous levels\end{tabular}}                                                                  & \vcell{Water}                         \\[-\rowheight]
\printcelltop                                                   & \printcelltop               & \printcelltop           & \printcelltop             & \printcelltop                                                                                                     & \printcelltop                                                             & \printcelltop                    \\ 
\hline
\vcell{2020}                                                    & \vcell{Texas, USA}                    & \vcell{\begin{tabular}[b]{@{}l@{}} Supply Chain\\Attack\end{tabular}}                & \vcell{Malicious Code}                  & \vcell{\begin{tabular}[b]{@{}l@{}} Network Management\\ Company and its clients\end{tabular}}                                                                                                           & \vcell{\begin{tabular}[b]{@{}l@{}} Compromised network \\and system data\end{tabular}}        & \vcell{IT}                 \\[-\rowheight]
\printcelltop                                                   & \printcelltop               & \printcelltop           & \printcelltop             & \printcelltop                                                                                                     & \printcelltop                                                             & \printcelltop                    \\ 
\hline
\vcell{2020}                                                    & \vcell{Japan}                    & \vcell{\begin{tabular}[b]{@{}l@{}} Unauthorized\\Access\end{tabular}}                & \vcell{Stolen Credentials
}                  & \vcell{\begin{tabular}[b]{@{}l@{}} Japan's largest electronic\\ equipment manufacturer\end{tabular}}                                                                                                          & \vcell{\begin{tabular}[b]{@{}l@{}} Compromised details \\of missile designs\end{tabular}}                                                                   & \vcell{Defense}                         \\[-\rowheight]
\printcelltop                                                   & \printcelltop               & \printcelltop           & \printcelltop             & \printcelltop                                                                                                     & \printcelltop                                                             & \printcelltop                    \\ 
\hline
\vcell{2018}                                                    & \vcell{New York, USA}         & \vcell{Possible botnets}          & \vcell{\begin{tabular}[b]{@{}l@{}}Demand \\Manipulation \end{tabular}}         & \vcell{Transformers}                       & \vcell{\begin{tabular}[b]{@{}l@{}}Transformer explosion \\and grounded flights \end{tabular}}                                                                   & \vcell{Energy}                   \\[-\rowheight]
\printcelltop                                                   & \printcelltop               & \printcelltop           & \printcelltop             & \printcelltop                                                                                                     & \printcelltop                                                             & \printcelltop                    \\ 
\hline
\vcell{2017}                                                    & \vcell{Middle East}         & \vcell{Triton}          & \vcell{Malware~~}         & \vcell{\begin{tabular}[b]{@{}l@{}}PLC Malware targeting \\plant safety system\end{tabular}}                       & \vcell{Forced shut down of the power plant}                                                                  & \vcell{Energy}                   \\[-\rowheight]
\printcelltop                                                   & \printcelltop               & \printcelltop           & \printcelltop             & \printcelltop                                                                                                     & \printcelltop                                                             & \printcelltop                    \\ 
\hline
\vcell{2016}                                                    & \vcell{Kiev, Ukraine}       & \vcell{Crash Override}  & \vcell{Malware}           & \vcell{\begin{tabular}[b]{@{}l@{}}Malware with 101/104 \\ IEC 61850 modules\end{tabular}}                            & \vcell{\begin{tabular}[b]{@{}l@{}}Electricity supply interruption\\ and delayed recovery operations\end{tabular}}                                                                  & \vcell{Energy}                   \\[-\rowheight]
\printcelltop                                                   & \printcelltop               & \printcelltop           & \printcelltop             & \printcelltop                                                                                                     & \printcelltop                                                             & \printcelltop                    \\ 
\hline
\vcell{2015}                                                    & \vcell{Kiev, Ukraine}       & \vcell{Black Energy}    & \vcell{FDI}               & \vcell{\begin{tabular}[b]{@{}l@{}}Control centre PCs \\remote-controlled, \\RTUs infected/destroyed\end{tabular}} & \vcell{\begin{tabular}[b]{@{}l@{}}Blackout 225k \\customers\end{tabular}} & \vcell{Energy}                   \\[-\rowheight]
\printcelltop                                                   & \printcelltop               & \printcelltop           & \printcelltop             & \printcelltop                                                                                                     & \printcelltop                                                             & \printcelltop                    \\ 
\hline
\vcell{2012}                                                    & 
\vcell{\begin{tabular}[b]{@{}l@{}} Saudi Arabia\\ and Qatar\end{tabular}}

& \vcell{Shamoon \cite{Bronk2013}}         & \vcell{Malware}           & \vcell{\begin{tabular}[b]{@{}l@{}}Malware affected \\generation/delivery\\~at Aramco and RasGass\end{tabular}}    &  \vcell{\begin{tabular}[b]{@{}l@{}} Disrupted Aramco Oil\\ Company for two weeks\end{tabular}}                                                                  & \vcell{Energy}                   \\[-\rowheight]
\printcelltop                                                   & \printcelltop               & \printcelltop           & \printcelltop             & \printcelltop                                                                                                     & \printcelltop                                                             & \printcelltop                    \\ 
\hline
\vcell{2010}                                                    & \vcell{Iran}                & \vcell{Stuxnet}         & \vcell{Malware}                  & \vcell{\begin{tabular}[b]{@{}l@{}}PLCs infected via \\Eng. PCs\end{tabular}}                                      & \vcell{1000 centrifuges}                                                  & \vcell{Energy}                   \\[-\rowheight]
\printcelltop                                                   & \printcelltop               & \printcelltop           & \printcelltop             & \printcelltop                                                                                                     & \printcelltop                                                             & \printcelltop                    \\ 
\hline
\vcell{2008}                                                    & \vcell{Turkey}              & \vcell{}                & \vcell{FDI}               & \vcell{\begin{tabular}[b]{@{}l@{}}Control system \\parameters~of\\~the oil pipeline\end{tabular}}                 & \vcell{Oil Explosion}                                                     & \vcell{Oil}                      \\[-\rowheight]
\printcelltop                                                   & \printcelltop               & \printcelltop           & \printcelltop             & \printcelltop                                                                                                     & \printcelltop                                                             & \printcelltop                    \\ 
\hline
\vcell{2007}                                                    & \vcell{Idaho, USA}          & \vcell{Aurora}          & \vcell{FDI}               & \vcell{\begin{tabular}[b]{@{}l@{}}Circuit Breaker\\~of a Generator\end{tabular}}                                  & \vcell{\begin{tabular}[b]{@{}l@{}}Generator \\Explosion\end{tabular}}     & \vcell{Energy}                   \\[-\rowheight]
\printcelltop                                                   & \printcelltop               & \printcelltop           & \printcelltop             & \printcelltop                                                                                                     & \printcelltop                                                             & \printcelltop                    \\ 
\hline
\vcell{2003}                                                    & \vcell{Ohio, USA}           & \vcell{Slammer Worm}    & \vcell{Malware}           & \vcell{\begin{tabular}[b]{@{}l@{}}Nuclear plant \\control system\end{tabular}}                                    & \vcell{\begin{tabular}[b]{@{}l@{}}System \\Display Off\end{tabular}}      & \vcell{Energy}                   \\[-\rowheight]
\printcelltop                                                   & \printcelltop               & \printcelltop           & \printcelltop             & \printcelltop                                                                                                     & \printcelltop                                                             & \printcelltop                    \\ 
\hline

\end{tabular}
\end{table*}



\section {Vulnerabilities and Attack Vectors in Smart Grid}

Generally, the attacks are divided into two categories: passive attacks and  active attacks. The passive refers to those that target the system information; however, it doesn't directly affect system resources. Examples of these attacks are spying, eavesdropping and traffic analysis. On the other hand, the active attacks aim to make changes to the data and information. These include malware attacks and DoS. Based on the reported incidents often the active attacks caused severe consequences. 
Several attack vectors have been addressed in the existing literature. Based on the very few available threat models \cite{9372271,9126843}, these could be aimed at the monitoring device level, substation level, or through remote access. At each of these levels, a variety of cyber vulnerabilities have been addressed. 
Among all the attacks, FDI has been identified as the most common class of threat and the most challenging with the widest impacts \cite{Musleh2020VulnerabilitiesOverview}. 
From \cite{Musleh2020} the smart grid structure is vulnerable to 4 major group of attacks (Figure 2): 

•	Physical based (e.g. damage to physical part, tampering with device input, Electromagnetic radiation, heat, light and sound emmision), 

•	Cyber based (e.g. changes to software or the firmware, Code or command manipulation),

•	Communication based (e.g. the physical link breakdown, GPS spoofing)  and, 

•	Network based (e.g. manipulate the data within packets in network, Denial of Service). 


FDI has been found the common type of attack in the all of the above 4 group of attacks. 

At the physical layer, FDI attack is feasible by manipulating the device input. The data can be manipulated at cyber layer at application level with no changes to the codes. Injecting a falsified data via GPS is a possible scenario of communication layer attack using FDI. When the data manipulation occurs in network packets it is also considered as FDI \cite{Musleh2020}.    



\begin{figure}[h]
    \includegraphics[width=\linewidth]{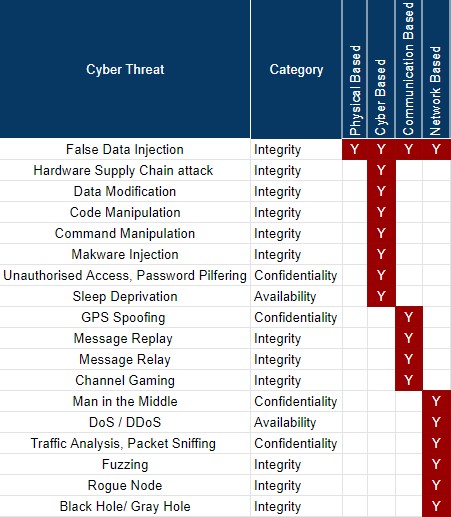}


  \caption{Common attacks in smart grid} \label{Common Attacks}
 
\end{figure}

\section {Transformers Attack Surface}

Figure 3 \cite{OmicronDetectingNetworks} shows the attack surface using a common substation transformer architecture. Remote access points (e.g. corporate office, IT, network control centre and the home internet) can be compromised by malicious actors. A common pathway is the loss of login credentials of diagnostic equipment using a spoofed page. 
Testing equipment for secondary systems or protection equipment testing is usually connected to the station bus. This creates another attack surface \cite{Klien,Klien2020DesignArchitecture}. The testing companies have realised  the risk and isolated the testing PC/laptop from the station bus, however, the test equipment still needs to remain  connected to the bus to allow for the testing to be completed.
Unauthorised access through the engineering PC can target the gateway and the IEDs by introducing a malicious configuration file.

Based on the very few threat models \cite{9372271, Olijnyk2022DesignTransformer}, generally the following common vulnerabilities and exposures are addressed: 

1)	Tampering

2)	Spoofing  

3)	Denial of Service

4)	Information Disclosure

5)	Repudiation

6)	Elevation of Privilege

These mainly target the monitoring equipment associated with transformers. When it comes to the transformer accessories such as the OLTC, FDI and hidden command are the main attacks discussed in \cite{10.1007/978-3-319-23802-9_20} and \cite{9126843}. FDI has been studied extensively by many researchers in smart grids \cite{TRAN2020DesignGeneration,Musleh2020,Xu2021DetectionLearning,Kumar2017EfficientGrids}, however, in the case of power transformers, there have been only a few studies considering the FDI. Papers \cite{Khaw2021ARelays} and \cite{Jahromi2021DataProtection} discuss the supply chain issue and the vulnerability of the MUs, as they are collecting sensitive data such as voltages and currents.
Paper \cite{Olijnyk2022DesignTransformer} explored the methods of attack when manipulating the physical data of the transformers. Researchers have covered the following:

•	Physics-centric attack,

•	Interposition attacks,

•	Coordinated distributed attacks,

•	Physics-centric malware emulation attack,

•	Harmonic restraint attack emulation,

•	Differential attack emulation,

•	OLTC/Tap-changer attack emulation.

Paper \cite{Automated2020ParisGrid} classified the process level data information as the highest risk and the most sensitive data. This is where the transformers are located and with the reference to \cite{Automated2020ParisGrid}, the next step is to consider mitigation strategies.

\begin{figure}[h]
    \includegraphics[width=\linewidth]{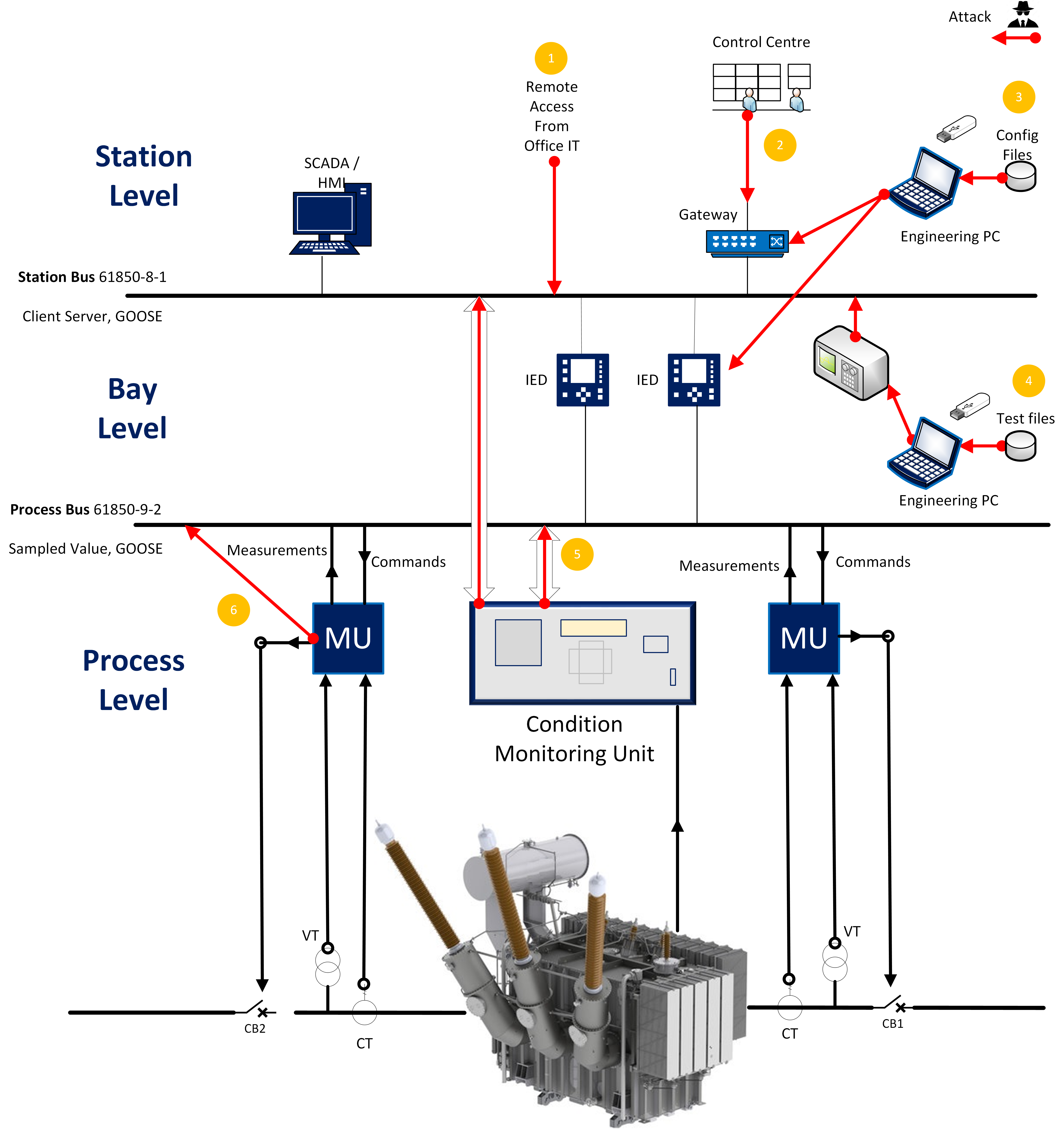}


  \caption{Substation architecture and attack vectors} \label{Attack Vectors}
 
\end{figure}

\section{Detection Algorithms}

Smart grid asset is proven to be vulnerable to data integrity attacks. The FDI attacks are targeting the power transformers. To tackle this issue detection algorithms have been developed over the years.
Detection algorithms could mainly categorised into two categories of rule-based methods and AI-based techniques.


\subheading{Rule-Based Methods.}
Paper \cite{Musleh2020} provides a summary of the largest sets of the algorithms used in the smart grid context for the mostly referred FDI attack. Figure 4 shows these algorithms can be classified into two categories; model based and data based. Figure 5 shows the statistics for algorithms used in FDI attacks.
From the statistical analysis, there is a growing appetite for the use of data-driven algorithms. This is a consideration for both researchers in academia and utilities globally \cite{Kulkarni2020CyberStudy,Sattinger2020CriticalThreats,Autonoma2020AnIntelligence}. And has come about mainly due to the limitation of the model-based algorithm in relation to the growing grid and its increased complexity. The recent literature \cite{Karimipour2019AGrids} suggests unsupervised deep learning can be computationally efficient and reliable for detection of attacks in a large scale smart grid. This results in high rates of detection (99\%) and are in good agreement when compared to the simulation results of an IEEE 2448 bus system. Another South Australian utility detailed some case studies and showed by example how machine learning techniques, such as deep learning, helped them with detecting and responding to cyber threats \cite{Sattinger2020CriticalThreats}.  

\begin{figure}
 \centering

    \includegraphics[width=\linewidth]{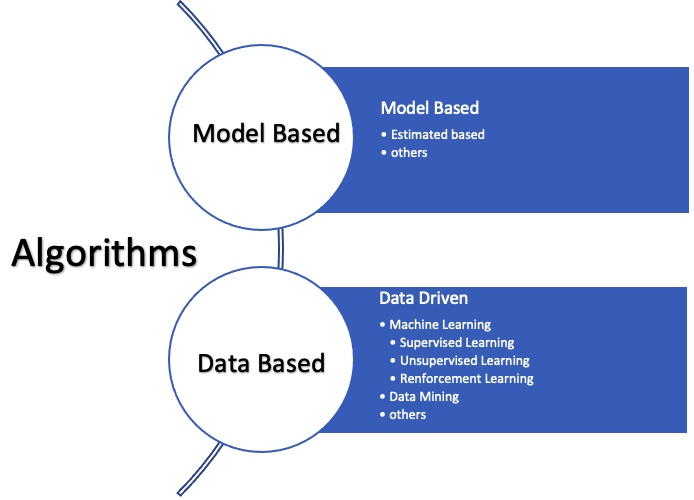}


  \caption{Algorithms classification}\label{Algorithms}
 
\end{figure}

 \begin{figure}
 \centering

    \includegraphics[width=\linewidth]{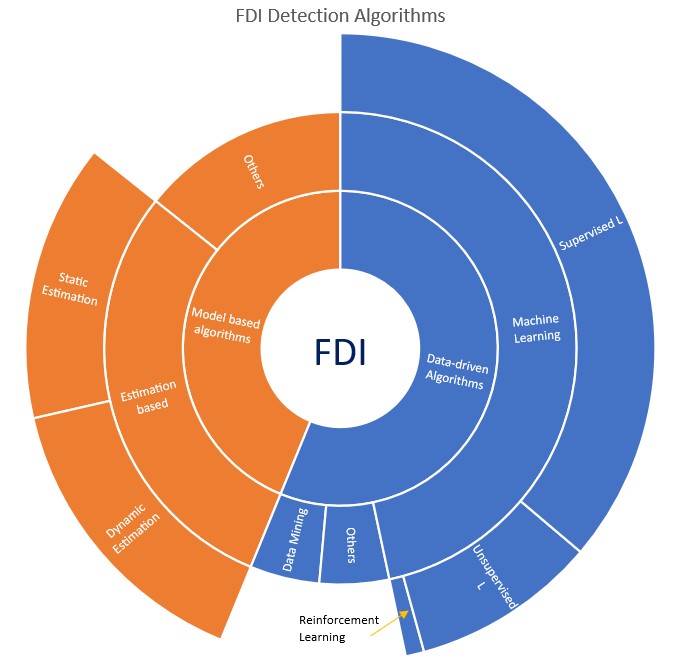}


  \caption{Statistics of used algorithms for FDI attack in existing literature.}\label{Algorithms Statistics}
 
\end{figure}




\subheading{Use of AI Techniques for Power
Transformers.}
To date, there has been little research on the disruption of smart grid operation due to targeted cyberattacks on power transformers. With the move to the next generation of so-called smart transformers, there is an increasing concern about their cybersecurity \cite{GoudCyberExplosions,SiemensEnergySensformerSecurity}, with only limited research towards the development of an AI-based cyber defence framework for their  protection.
Some previously used methods \cite{Karimipour2019AGrids} are not well suited to large and complex electrical networks due to their high computation demands and system storage requirements. The focus of the existing literature has been on network security~\cite{zheng2020towards,janicke2020security}. 
To tackle smart grid security threats for smart transformers, suitable models are required using source classification and simulation tools for threat sensing and response. The focus needs to be on technical operation security and the system’s data management security. This includes the security in reliability and resiliency of operations and real-time recording, monitoring and storing of data and information.  
As the present sensors and accessories on transformers have limited computational power, adversarial attacks with a focus on poisoning attacks needs to be explored. A novel threat profile for these smart components is still missing from the existing literature. 
Identifying the different sources of security threats and challenges, including the known and unknown attacks is to be studied and identified. These are to be classified, and the application of suitable models, analysis, simulation, and optimization solutions are to be reviewed.
There has been extensive research focused on smart meters, to identify various attacks and defences against them. However, there is limited work to automatically defend the smart transformers. The nature of periodical data collection is completely unique and different from smart meters. This includes real-time monitoring of transformer vital parameters (Voltage regulation and OLTC monitoring, temperature and cooling system control, online Dissolved Gas Analysis, online bushing health information, etc.).
Machine learning algorithms such as clustering algorithms, regression or classification are still to be explored specifically for transformers when it is required to distinguish between real data and data from bad actors.
Previous studies have shown device attacks, application service attacks, network attacks, web interface attacks and data integrity attacks are all applicable to the smart meter context. The practicalities of this attack on smart transformers are to be explored. Many of the existing AI based algorithms are built around a particular understanding of historical data. However, the attack surface and the nature of attacks are evolving. Therefore, the use of AI based frameworks, making use of machine learning techniques appear to offer an effective way to stay ahead of cybersecurity.

\section{Conclusion}
Power transformers are complex elements of the power system which have their own expertise in operation and assessment. Their concentrated nature, high cost and long lead time, make them a high priority target for malicious actors.  In the past they have been relatively safe from external interference but with the advent networked controls and monitoring they are now vulnerable. Studies have shown there is a potential for cyber threats to the important components of power transformers, including OLTC, DGA systems, connected CBs, differential protection and phase shifting transformers.  A new generation of intelligent monitoring systems that are integral to a Smart Transformer are coming into wider use. As these technologies are evolving and the attackers are motivated, it is suggested that AI and machine learning approaches may be necessary to best secure such systems from not only network-based attacks but also malicious internal actors.
AI diagnostic systems, in particular those employing  state estimation based algorithms, are best place at present to block a vast range of attacks from external and internal threats. To counter evolution in threats, it is important to detect stealthy attacks which appear to need a degree of unsupervised learning approaches. Presently there is no mitigation strategy in the existing literature for online diagnostic systems. Transformer differential protection has received more attention,  however, the suggested mitigation algorithms have been only tested against a very limited number of hypothetical scenarios and hence their robustness against advanced cybersecurity is unclear. 
Due to the higher rates of attack and resources available to counter them, USA utilities are leading in this field.  Australian utilities are also learning from their own and global experience and actively working to keep their networks secure. 

This research summarises the cybersecurity challenges associated with power transformers within electrical networks and brings together an understanding of attack vectors for transformer components and their near associated devices. As a new generation of power transformers equipped with smart monitoring system enters use, the issues highlighted are a pathway to develop transformer specific mitigation strategies. The studied literature provided some insight to the cyber protection of smart transformers, however, there is still much to be explored. With continuous advancements in cybersecurity, there is a need for equally continuously development in robust counter measures.     
It is encouraged to investigate the provided mitigation strategies such as the one in paper \cite{Jahromi2020} with a larger range of attack scenarios to provide a better line of defence for these assets . The complex, dynamic and ever increasing nature of the grid along with the various states of the equipment and their operating conditions needs to be considered \cite{Transgrid2021EnergyVision,Musleh2020}. Another problem for study is how best to distinguish an attackers telemetry indications and operating commands from the true state of the transformer and its associated systems, which has been touched on for a Wide-Area Monitoring Systems in \cite{Musleh2021OnlineSystems}. 
   



%






%





\ifCLASSOPTIONcaptionsoff
  \newpage
\fi




{\footnotesize	

\bibliographystyle{IEEEtran}
\bibliography{IEEEabrv,Bibliography}
}
 \begin{IEEEbiography}[{\includegraphics[width=1in,height=1.25in,clip,keepaspectratio]{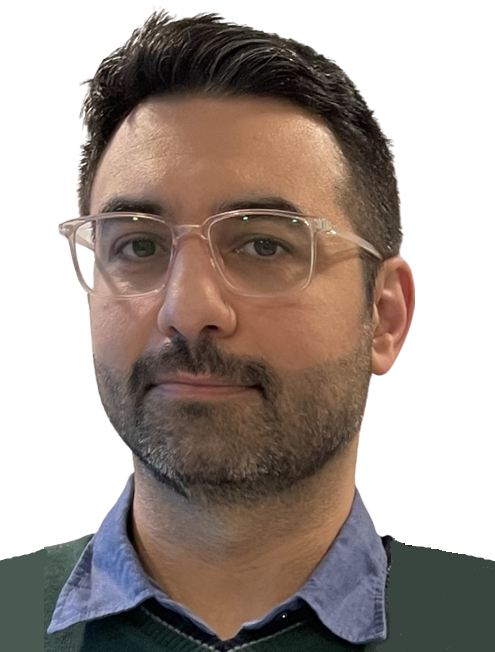}}]{Hossein Rahimpour}
 received the M.Ph.E.E. degree from the University of Newcastle, Australia, in 2018, and currently he is pursuing his Ph.D. degree at the University of UNSW. From 2006 to 2021, he was a Senior Engineer in power industry, where he was involved with high voltage systems. His current research interests include cybersecurity, machine learning, smart grid, power transformers, high voltage. He is currently jointly supervised by UNSW and CSIRO's Data61 working towards his PhD.
 \end{IEEEbiography}
  \begin{IEEEbiography}[{\includegraphics[width=1in,height=1.25in,clip,keepaspectratio]{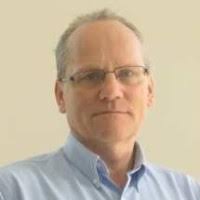}}]{Joe Tusek}
 received the B.E. degree from the University of Newcastle and the MBA degree from the Deakin University. He currently works at Verico. He does research in electrical plant condition assessment as well as instrumentation and test equipment design for industrial environments. Current research interest is related to improving the reliability of Frequency Response Analysis measurements in large power transformers.
 \end{IEEEbiography}
   \begin{IEEEbiography}[{\includegraphics[width=1in,height=1.25in,clip,keepaspectratio]{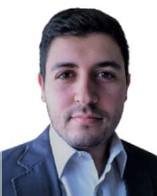}}]{Ahmed S. Musleh}
(Member, IEEE) received the M.Sc. degree in Electrical Engineering from the Petroleum Institute (currently Khalifa University), Abu Dhabi, UAE in 2016. He received the Abu Dhabi University Overall Award of
Excellence and the Petroleum Institute Graduate Fellowship in 2014 and 2015, respectively. Currently, he is pursuing his Ph.D. degree at the School of Electrical Engineering and Telecommunications, University of New South Wales, Sydney, Australia. His research interests include smart grid technologies, wide-area monitoring and control, cyber-physical security, and machine learning applications.
 \end{IEEEbiography}
   \begin{IEEEbiography}[{\includegraphics[width=1in,height=1.25in,clip,keepaspectratio]{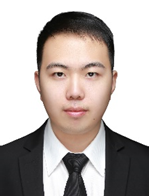}}]{Boyu Liu}
 (Member, IEEE) received the B.Eng. degree in electrical engineering and its automation from Xi’an University of Technology, Xi’an, China in 2017, and the M.Eng. degree in electrical engineering from the Northeastern University, Shenyang, China in 2021. Currently, he is a casual academic and pursuing his Ph.D. degree at the University of New South Wales, Sydney, Australia. His research interests include optimization in energy systems, convex optimization, evolutionary computation, and desalination and hydrogen applications in energy systems.
 \end{IEEEbiography}
 \begin{IEEEbiography}[{\includegraphics[width=1in,height=1.25in,clip,keepaspectratio]{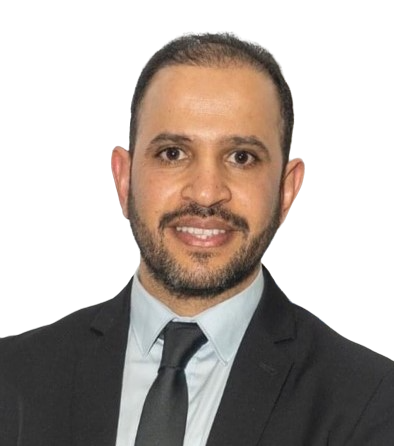}}]{Sharif Abuadbba}
 is a Senior Research Scientist at CSIRO’s Data61 and cybersecurity CRC fellow. Sharif  has  a  Ph.D.  in  computer  security  from RMIT  University,  Australia,  2017.  He  also  has several years of experience working as a senior research engineer with Californian-based technology companies. He has contributions to a few US IP filled Patents in cybersecurity. His specialist and interests include AI and cybersecurity, IoT-Cloud Security, System security, and watermarking.
 \end{IEEEbiography}
 \begin{IEEEbiography}[{\includegraphics[width=1in,height=1.25in,clip,keepaspectratio]{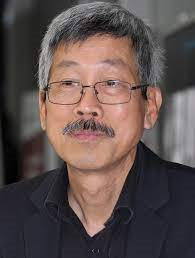}}]{Toan Phung}
(Senior Member, IEEE) received a Ph.D. in electrical engineering in 1998 and is currently an Associate Professor in the School of Electrical Engineering and Telecommunications at the University of New South Wales, Sydney, Australia. He has over 30 years of practical research/development experience in partial discharge measurement, and in on-line condition monitoring of high-voltage equipment. His research interests include electrical insulation (dielectric materials and diagnostic methods), high-voltage engineering (generation, testing and measurement techniques), electromagnetic transients in power systems, and power system equipment (design and condition monitoring methods).(Based on document published on 5 December 2018). 

 \end{IEEEbiography}
\begin{IEEEbiography}[{\includegraphics[width=1in,height=1.25in,clip,keepaspectratio]{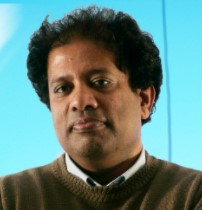}}]{Aruna Seneviratne}
 (Senior Member, IEEE) is currently a Foundation Professor of telecommunications with the University of New South Wales, Australia, where he holds the Mahanakorn Chair of telecommunications. He has also worked at a number of other Universities in Australia, U.K., and France, and industrial organizations, including Muirhead, Standard Telecommunication Labs, Avaya Labs, and Telecom Australia (Telstra). In addition, he has held visiting appointments at INRIA, France. His current research interests are in physical analytics: technologies that enable applications to interact intelligently and securely with their environment in real time. Most recently, his team has been working on using these technologies in behavioral biometrics, optimizing the performance of wearables, and the IoT system verification. He has been awarded a number of fellowships, including one at British Telecom and one at Telecom Australia Research Labs.
 \end{IEEEbiography}





\vfill


\end{document}